\documentclass[twoside,twocolumn,9pt]{article}
\usepackage{extsizes}
\usepackage[super,sort&compress,comma]{natbib} 
\usepackage[version=3]{mhchem}
\usepackage[left=1.5cm, right=1.5cm, top=1.785cm, bottom=2.0cm]{geometry}
\usepackage{balance}
\usepackage{mathptmx}
\usepackage{sectsty}
\usepackage{graphicx} 
\usepackage{lastpage}
\usepackage[format=plain,justification=justified,singlelinecheck=false,font={stretch=1.125,small,sf},labelfont=bf,labelsep=space]{caption}
\usepackage{float}
\usepackage{fancyhdr}
\usepackage{fnpos}
\usepackage[english]{babel}
\addto{\captionsenglish}{%
  
}
\usepackage{array}
\usepackage{droidsans}
\usepackage{charter}
\usepackage[T1]{fontenc}
\usepackage[usenames,dvipsnames]{xcolor}
\usepackage{setspace}
\usepackage[compact]{titlesec}
\usepackage{hyperref}

\usepackage{epstopdf}

\definecolor{cream}{RGB}{222,217,201}

\usepackage{color}
\newcommand{\Q}[1]{{{\textcolor{black}{#1}}}}
\usepackage{amsmath,amssymb}
\usepackage{upgreek}
\usepackage{changes}

\AppendGraphicsExtensions{.tif}

\begin{document}

\pagestyle{fancy}
\thispagestyle{plain}
\fancypagestyle{plain}{
\renewcommand{\headrulewidth}{0pt}
}

\makeFNbottom
\makeatletter
\renewcommand\LARGE{\@setfontsize\LARGE{15pt}{17}}
\renewcommand\Large{\@setfontsize\Large{12pt}{14}}
\renewcommand\large{\@setfontsize\large{10pt}{12}}
\renewcommand\footnotesize{\@setfontsize\footnotesize{7pt}{10}}
\makeatother

\renewcommand{\thefootnote}{\fnsymbol{footnote}}
\renewcommand\footnoterule{\vspace*{1pt}%
\color{cream}\hrule width 3.5in height 0.4pt \color{black}\vspace*{5pt}} 
\setcounter{secnumdepth}{5}

\makeatletter 
\renewcommand\@biblabel[1]{#1}            
\renewcommand\@makefntext[1]%
{\noindent\makebox[0pt][r]{\@thefnmark\,}#1}
\makeatother 
\renewcommand{\figurename}{\small{Fig.}~}
\sectionfont{\sffamily\Large}
\subsectionfont{\normalsize}
\subsubsectionfont{\bf}
\setstretch{1.125} 
\setlength{\skip\footins}{0.8cm}
\setlength{\footnotesep}{0.25cm}
\setlength{\jot}{10pt}
\titlespacing*{\section}{0pt}{4pt}{4pt}
\titlespacing*{\subsection}{0pt}{15pt}{1pt}

\fancyfoot{}
\fancyfoot[LO,RE]{\vspace{-7.1pt}\includegraphics[height=9pt]{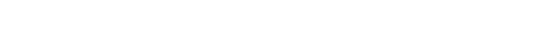}}
\fancyfoot[CO]{\vspace{-7.1pt}\hspace{11.9cm}\includegraphics{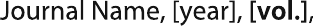}}
\fancyfoot[CE]{\vspace{-7.2pt}\hspace{-13.2cm}\includegraphics{head_foot/RF}}
\fancyfoot[RO]{\footnotesize{\sffamily{1--\pageref{LastPage} ~\textbar  \hspace{2pt}\thepage}}}
\fancyfoot[LE]{\footnotesize{\sffamily{\thepage~\textbar\hspace{4.65cm} 1--\pageref{LastPage}}}}
\fancyhead{}
\renewcommand{\headrulewidth}{0pt} 
\renewcommand{\footrulewidth}{0pt}
\setlength{\arrayrulewidth}{1pt}
\setlength{\columnsep}{6.5mm}
\setlength\bibsep{1pt}

\makeatletter 
\newlength{\figrulesep} 
\setlength{\figrulesep}{0.5\textfloatsep} 

\newcommand{\topfigrule}{\vspace*{-1pt}%
\noindent{\color{cream}\rule[-\figrulesep]{\columnwidth}{1.5pt}} }

\newcommand{\botfigrule}{\vspace*{-2pt}%
\noindent{\color{cream}\rule[\figrulesep]{\columnwidth}{1.5pt}} }

\newcommand{\dblfigrule}{\vspace*{-1pt}%
\noindent{\color{cream}\rule[-\figrulesep]{\textwidth}{1.5pt}} }

\makeatother

\twocolumn[
  \begin{@twocolumnfalse}
\vspace{1em}
\sffamily
\begin{tabular}{m{4.5cm} p{13.5cm} }

\includegraphics{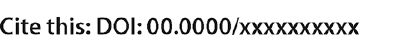} & \noindent\LARGE{\textbf{Protein solutions close to liquid--liquid phase separation 
exhibit a  universal osmotic equation of state and dynamical behavior
}} \\
\vspace{0.3cm} & \vspace{0.3cm} \\

 & \noindent\large{Jan Hansen,\textit{$^{a}$} 
 Stefan U. Egelhaaf,\textit{$^{a}$} and
Florian Platten,$^{\ast}$\textit{$^{a}$}}  \\

\includegraphics{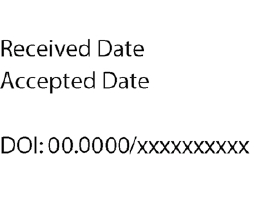} & \noindent\normalsize{
Liquid--liquid phase separation (LLPS) of protein solutions is governed by highly complex protein--protein interactions.
Nevertheless, it has been suggested that based on the extended law of corresponding states (ELCS), as proposed for colloids with short-range attractions, one can rationalize not only the thermodynamics, but also the structure and dynamics of such systems.
This claim is systematically and comprehensively tested here by static and dynamic light scattering experiments.
Spinodal lines, the isothermal osmotic compressibility $\kappa_\text{T}$ and the relaxation rate of concentration fluctuations $\Gamma$ are determined for protein solutions in the vicinity of LLPS.
All these quantities are found to exhibit a corresponding-states behavior.
This means that, for different solution conditions, these quantities are essentially the same if considered at similar reduced temperature or second virial coefficient.
For moderately concentrated solutions, the volume fraction $\phi$ dependence of $\kappa_\text{T}$ and $\Gamma$ can be consistently described by Baxter's model of adhesive hard spheres.
The off-critical, asymptotic $T$ behavior of $\kappa_\text{T}$ and $\Gamma$ close to LLPS is consistent with the scaling laws predicted by mean-field theory. 
Thus, the present work aims at a comprehensive experimental test of the applicability of the ELCS to structural and dynamical properties of concentrated protein solutions.
} \\
\end{tabular}
 \end{@twocolumnfalse} \vspace{0.6cm}  ]

\renewcommand*\rmdefault{bch}\normalfont\upshape
\rmfamily
\section*{}
\vspace{-1cm}


\footnotetext{\textit{$^{a}$~Condensed Matter Physics Laboratory, Heinrich Heine University, Universit\"atsstraße~1, 40225 D\"usseldorf, Germany; E-mail: florian.platten@hhu.de}.}






\section{Introduction}

The interactions between protein molecules in solution, in general, comprise repulsive and attractive contributions. If short-range attractions prevail, protein condensation is favorable.\cite{Gunton2007,Piazza2004,Vekilov2010,Zhang2014,McManus2016,Stradner2020}
If the attractions are strong, proteins are likely to form amorphous aggregates. If they are moderate, proteins are prone to crystallize or phase separate.
Protein solutions can form two coexisting liquid phases, one depleted and one enriched in proteins; this process is called liquid-liquid phase separation (LLPS).\cite{Muschol1997,Broide1996}
It has recently attracted considerable attention due to its relevance in cell biology, medicine, pharmaceutical industry, and protein crystallography, as illustrated by the following examples.
Membraneless organelles formed via LLPS in the cytosol represent a way of intracellular organization and can regulate biochemical reactions.\cite{Brangwynne2015,Alberti2021}
In vivo, LLPS can induce pathological protein aggregation, e.g., in the context of amyloid diseases, cataract and sickle-cell anemia.\cite{Pande2001,Galkin2002,Babinchak2020}
In biopharmaceuticals, LLPS is likely to increase solution viscosity and might raise immunogenicity.\cite{Salinas2010,Raut2016} Hence, conditions favoring LLPS should be avoided in formulation development. 
However, X-ray crystallographers can exploit conditions close to LLPS, as they might favor the growth of high-quality crystals needed for diffraction experiments.\cite{Vliegenthart2000,Galkin2000,Zhang2012a,Zhang2012b} 

The protein--protein interactions that govern the phase separation processes are highly complex. 
They are determined by the specific amino acid composition of the proteins, their internal molecular organization, their overall molecular shape with a heterogeneous distribution of functionally different moieties on their surfaces as well as on the physicochemical solution conditions. 
It is therefore challenging to decipher a physical rationale of protein phase behavior and the underlying interactions.

Nevertheless, coarse-grained approaches inspired by soft-matter physics have been applied to the highly complex interactions between protein molecules and have proven helpful in understanding the observed phase behavior.\cite{Stradner2020}
For example, the DLVO theory has been applied to describe the effects of salts, solvents and pH on the interactions of protein solutions under conditions relevant for crystallization and LLPS.\cite{Poon2000,Sedgwick2007,Pellicane2012,Goegelein2012,Kumar2019,Hansen2021b}
\Q{Colloid models have also been applied to describe the experimentally obtained structure factor of protein solutions}.\cite{Tardieu1999,Liu2005,Zhang2007,Chinchalikar2013,Wolf2014,Braun2017,Kundu2018}
With respect to LLPS of proteins, it would be appealing to apply the extended law of corresponding states (ELCS) and hence reduce the complexity of the situation.
For colloids with short-range attractions, the ELCS, as proposed by Noro and Frenkel,\cite{Noro2000} suggests that thermodynamic properties of the solution, such as the equation of state (EOS), are insensitive to the molecular details of the underlying interactions.
\Q{This has been tested by theory and simulations for simple colloidal model systems.\cite{Valadez-Perez2012}
The EOS is} governed by only few parameters, such as the second virial coefficient $B_2$, which can be considered as a measure of the strength of the interactions.  
For a spherosymmetric potential $U(r)$ with center-to-center distance $r$, its definition reads
\begin{equation}\label{eq:b}
B_2 = 2 \pi \int_0^\infty \left( 1 - \exp{\left[-\frac{U(r)}{k_\text{B} T} \right]} \right) r^2 \text{d}r \, 
\end{equation}
with thermal energy $k_\text{B}T$. 
Often, $B_2$ is normalized by the second virial coefficient of a corresponding hard-sphere system with the same diameter $\sigma$, $B_2^\text{HS}=(2\pi/3)\,\sigma^3$, and reported as  $b_2=B_2/B_2^\text{HS}$.
Theoretical and simulation studies of colloidal systems further suggest that the ELCS could also govern structural and dynamical properties of the solution.\cite{Foffi2005,Lu2008,Zaccarelli2008,Valadez-Perez2018}

Inspired by the ELCS as proposed for colloidal model systems, it has been found that also the LLPS binodals of protein solutions collapse onto a master curve if plotted in the $b_2$ vs. volume fraction $\phi$ plane and the repulsions of the different systems are alike or their differences are accounted for in terms of an effective $\sigma$.\cite{Gibaud2011,Platten2015,Platten2016,Bucciarelli2016,Hansen2022b}
Nevertheless, despite the importance of LLPS and the possibility to rationalize the underlying interactions by the ELCS, systematic and comprehensive studies on the applicability of the ELCS to protein solutions are scarce, especially with respect to the structural and dynamical properties of concentrated solutions.

Here, 
the osmotic equation of state and the collective dynamics of protein solutions under conditions close to LLPS are systematically investigated.
Static and dynamic light scattering experiments (SLS and DLS) are used to determine the isothermal osmotic compressibility $\kappa_\text{T}$ and the relaxation rate of concentration fluctuations $\Gamma$, respectively.
Lysozyme in brine is used as a model system for proteins with short-range attractions,\cite{Sedgwick2005} whose strength can be modulated by additives.\cite{Hansen2016}
We thus exploit that, for this system,
state diagrams (LLPS binodals) and interactions parameters ($b_2$)  have been reported previously.\cite{Hansen2022b}
The spinodal lines, as inferred from SLS, as well as $\kappa_\text{T}$ and $\Gamma$ are found to exhibit a corresponding-states behavior.
For moderately concentrated solutions, the $\phi$ dependence of $\kappa_\text{T}$ and $\Gamma$ can be consistently described by Baxter's model of adhesive hard spheres.
The off-critical, asymptotic $T$ behavior of $\kappa_\text{T}$ and $\Gamma$ close to LLPS is consistent with the scaling laws predicted by mean-field theory. 
Thus, the present work aims at a comprehensive experimental test of the applicability of the ELCS to structural and dynamical properties of concentrated protein solutions.

\section{Experimental methods} 

\subsection{Sample preparation}

Purified, salt-free hen egg-white lysozyme powder (Worthington
Biochemicals, cat.no. LS002933)), sodium chloride (NaCl; Fisher Chemicals), guanidine hydrochloride (GuHCl; Sigma, prod.~no.~ G4505) and sodium acetate (NaAc; Merck, prod.~no.~ 1.06268) were used without further purification.

Ultrapure water with a minimum resistivity of 18~M$\Omega$cm was used to prepare buffer and salt stock solutions.
They were filtered several times meticulously (nylon membrane, pore size $0.2~\upmu\mathrm{m}$) in order to remove dust particles.
The concentration of salt stock solutions, prepared in buffer, was determined from the excess refractive index.
The protein powder was dissolved in a 50~mM NaAc buffer solution.
The solution $p$H was adjusted to $p$H 4.5 by adding small amounts of hydrochloric acid. 
At this $p$H value, lysozyme carries a positive net charge $Q = + 11.4~e$,\cite{Tanford1972} where $e$ is the elementary charge. 
Solutions with an initial protein concentration $c \approx 40-70$~mg/mL were passed several times through an Acrodisc syringe filter with low protein binding (pore size $0.1~\upmu\mathrm{m}$; Pall, prod.~no.~ 4611) in order to remove impurities and undissolved proteins. 
Then, the protein solution was concentrated by a factor of $4-7$ using a stirred ultrafiltration cell (Amicon, Millipore, prod.~no.~ 5121) with an Omega 10 k membrane disc filter (Pall, prod.~no.~ OM010025). 
The retentate was used as concentrated protein stock solution.
Its protein concentration was determined by UV absorption spectroscopy or refractometry.\cite{Platten2015b} 
Protein concentrations $c$ are related to the protein volume fraction $\phi = c \, v_\text{p}$, where $v_\text{p} = 0.740$~mL/g is the specific volume of lysozyme, as inferred from densitometry.\cite{Platten2015b} 
Sample preparation was performed at room temperature $(21\pm2)~^\circ\mathrm{C}$.
Samples were prepared by mixing appropriate amounts of buffer, protein and salt stock solutions.
Samples with cloud-points close to or above room temperature were slightly preheated to avoid (partial) phase separation upon mixing.

\subsection{Static light scattering (SLS) experiments}

Light scattering experiments\cite{Berne1976}  
were performed with a 3D light scattering apparatus (LS Instruments AG) with a wavelength $\lambda =632.8$~nm.

Samples with $10~\text{mg/mL} \leq c \leq 160~\text{mg/mL}$ ($0.007 \leq \phi \leq 0.12$) were investigated at selected temperatures $12.0~^\circ\mathrm{C} \leq T \leq 43.0~^\circ\mathrm{C}$. 
\Q{As the samples are metastable with respect to crystallization, samples are analyzed directly after preparation.}
Samples investigated at different $T$ were prepared separately.
\Q{In order to minimize spurious effects of any residual dust or undissolved aggregates that were not removed by filtration,}
the samples were filled into thoroughly cleaned cylindrical glass cuvettes (diameter 10~mm) and centrifuged (Hettich Rotofix 32A) 30~min at typically 2,500~g prior to the measurements.
They were then very carefully placed into the temperature-controlled vat of the instrument filled with decalin.

The time-averaged scattered intensity was recorded. 
In order to check sample quality, also DLS experiments (see below) were performed on the same samples. 
Samples with indications of aggregates or dust particles were discarded.
The meticuluos filtration \Q{and centrifugation} procedure was essential to obtain reproducible light scattering data.
However, this protocol did not work for samples with $c > 160~\text{mg/mL}$, possibly due to aggregation or crystallization\cite{Sedgwick2005,Hentschel2021}.

The absolute scattering intensity, i.e.~the excess Rayleigh ratio $R$, varies with protein concentration $c$ and temperature $T$. 
It was determined using toluene as a reference according to
\begin{eqnarray}
R(c,T) = \frac{\left\langle I_\text{p} (c,T) \right\rangle - \left\langle I_\text{s} (T) \right\rangle}{\left\langle I_\text{t} (T) \right\rangle - \left\langle I_\text{dc} \right\rangle} \, \frac{n(c,T)^2}{n_\text{t}(T)^2} \, R_\text{t}(T)
\end{eqnarray}
with the time-averaged scattered intensities of sample, solvent, toluene, and background (blocked beam), $\left\langle I_\text{p} \right\rangle$, $\left\langle I_\text{s} \right\rangle$, $\left\langle I_\text{t} \right\rangle$, and $\left\langle I_\text{dc} \right\rangle$, respectively, 
the refractive indices of the sample and toluene, $n$ and $n_\text{t}$, and
the Rayleigh ratio of toluene at the measurement temperature and wavelength, $R_\text{t}$.
The values of $n_\text{t}$\cite{Keefe2005} and $R_\text{t}$\cite{Lundberg1964,Bender1986,Narayanan2003,Itakura2006} were inferred from literature data.
The temperature dependence of $R_\text{t}$ was determined from the temperature dependence of the intensity scattered by a toluene sample.\cite{Goegelein2012}

The refractive index of a sample solution, $n$, was measured with a temperature-controlled Abbe refractometer (Model 60L/R, Bellingham \& Stanley) operated with a HeNe laser ($\lambda=632.8~\mathrm{nm}$) and at the temperature of the SLS experiment. Refractive index increments, $\text{d}n/\text{d}c$, were obtained from linear fits to the dependence of $n$ on $c$. 

In one-component solutions, the excess scattering typically contains information on the shape and size of the particles as well as the particle arrangement.
Their contributions are reflected in the form factor $P(Q)$ and the structure factor $S(Q)$, respectively,  
where $Q = (4 \pi n / \lambda) \sin{(\theta/2)}$ is the magnitude of the scattering vector with the scattering angle $\theta$.
Then, the excess Rayleigh ratio $R$ reads:
\begin{eqnarray}\label{eq:rkc}
R(Q) = K\, c \, M \, P(Q) \, S(Q)
\end{eqnarray}
with the average molar mass of the particle $M$ (with $M = 14 320~\text{g/mol}$ for lysozyme) and an optical constant $K$ given by
\begin{eqnarray}
K(T) = \frac{4 \pi^2 n_\text{s}(T)^2}{N_\text{A} \lambda^4} \left( \frac{\text{d}n}{\text{d}c}(T) \right)^2 
\end{eqnarray}
with the refractive index of the solvent $n_\text{s}$ and Avogadro's number $N_\text{A}$.

The effective protein diameter $\sigma=3.4~\text{nm}$ is small compared to $\lambda$, implying $\sigma \, Q \ll 1$.
Nevertheless, for selected samples, experiments were performed at $30^\circ \leq \theta \leq 150^\circ$ and
$R$ was indeed found to be independent of $\theta$ and hence of $Q$.
In most of our experiments thus only one angle $\theta=90^\circ$ and hence $Q\approx 0.018~\text{nm}^{-1}$ was investigated.
In addition to the DLS experiments, 
the independence of $R$ on $\theta$
also suggested that large particles, such as impurities or aggregates, were absent.

Moreover, in the low-$Q$ limit,
\begin{eqnarray}
S(Q\to 0)= \frac{\kappa_\mathrm{T}}{\kappa_\mathrm{T}^\mathrm{id}}
\end{eqnarray}
with the isothermal osmotic compressibility of the sample and that of an ideal solution, $\kappa_\mathrm{T}$ and $\kappa_\mathrm{T}^\mathrm{id}$, respectively.\cite{Hansen2006}
The latter is given by $\kappa_\mathrm{T}^\mathrm{id}=(k_\text{B}T\,\rho)^{-1}$
with particle number density $\rho$.

\subsection{Dynamic light scattering (DLS) experiments}

\Q{The samples investigated by SLS were also studied by DLS.}
In the vicinity of the LLPS spinodal or critical point,
multiple scattering is expected to be important.
In order to suppress these contributions,
the 3D cross-correlation technique\cite{Urban1998} is applied here.
The measured cross-correlation function $g_2(\Delta t)$ depends on lag time $\Delta t$ and is defined as
\begin{eqnarray}
g_2(\Delta t) = \frac{\langle I_1(t) I_2(t+\Delta t) \rangle}{\langle I_1(t)\rangle \langle I_2(t) \rangle}
\end{eqnarray}
with averages over time $t$ denoted by angular brackets and
the intensities measured by detector 1 and 2, $I_1$ and $I_2$, respectively. 
From $g_2(\Delta t)$, the intermediate scattering function (ISF) $f(\Delta t)$ is calculated via the Siegert relation:\cite{Frisken2001}
\begin{eqnarray}\label{eq:sieg}
g_2(\Delta t) = B + \beta |f(\Delta t)|^2
\end{eqnarray}
with the baseline factor $B$ which usually equals 1 and the intercept factor $\beta$ which depends on the detection optics. For 3D cross-correlation, $\beta \leq 0.25$.\cite{Schaetzel1991} 
The ISFs were analyzed by an exponential decay using a second-order cumulant ansatz:\cite{Koppel1972,Frisken2001,Mailer2015}
\begin{eqnarray}\label{eq:cum}
f(\Delta t) = \exp{\left( - \Gamma \, \Delta t\,  \left( 1 + \frac{1}{2} \mu_2 (\Delta t)^2 \right) \right)}
\end{eqnarray}
with the average relaxation rate related to local concentration fluctuations $\Gamma$ and \Q{the second cumulant $\mu_2$.
The normalized standard deviation $\sqrt{\mu_2}/\Gamma$ is a measure of the departure from a single exponential decay; here}, in many cases $\sqrt{\mu_2}/\Gamma < 0.04$ and in some larger \Q{(typically 0.3)}.
As the ISFs 
show a single-exponential decay and hence indicate
diffusive behavior,
the relaxation rate can be related to the collective diffusion coefficient $D_\text{c}$ of the local concentration fluctuations via
$\Gamma = D_\text{c} Q^2$.

\subsection{Optical microscopy}

For selected samples, the microscopic morphologies of the condensed protein states
as well as the phase separation kinetics (nucleation, domain formation or coarsening) were monitored by conventional optical microscopy.
Sample solutions were prepared at a temperature above the cloud-point, typically at $30~^\circ\mathrm{C}$, and filled into a capillary.\cite{Jenkins2008}
The capillary was mounted onto a home-built temperature-controlled microscope stage equipped with a thermoelectric cooler.
The samples were quenched with about $0.5~\mathrm{K/s}$ to specific temperatures and observed using an inverted brightfield microscope (Nikon Eclipse Ti-2) equipped with a $10\times$ plan-fluor objective (Nikon) and a CMOS camera (Allied Vision, Mako U-130B, $512 \times 512~\mathrm{px}^2$) for at least $2~\mathrm{h}$. 
The pixel pitch was $0.485~\upmu\mathrm{m}/\mathrm{px}$.

\section{Results and Discussion}

\begin{figure}[h!]
\includegraphics[width=\columnwidth]{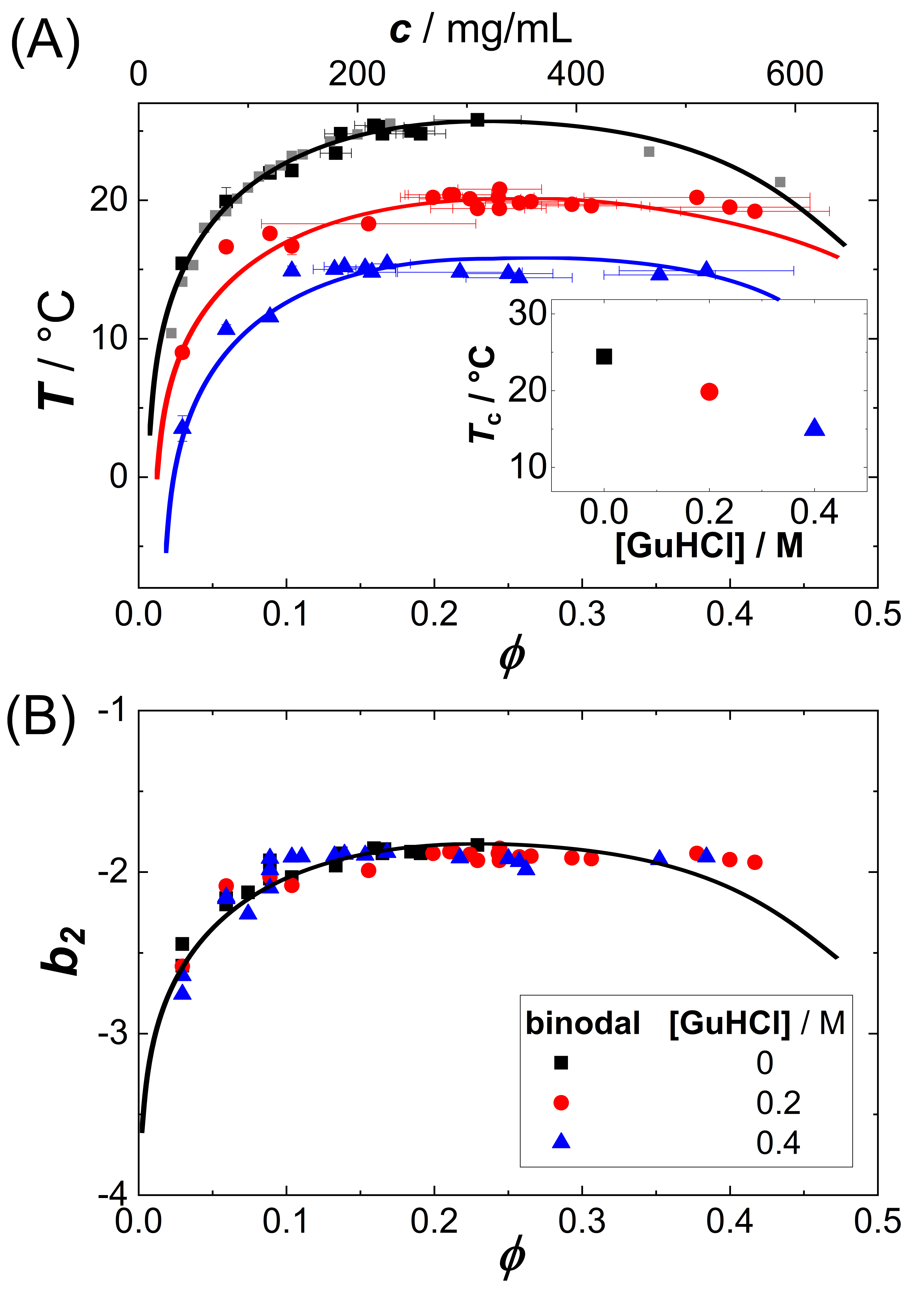} 
  \caption{
  Liquid--liquid phase separation of protein solutions: 
 (A)  Binodals of lysozyme solutions (pH 4.5, 0.9~M NaCl) in the temperature $T$ vs. volume fraction $\phi$ (or protein concentration $c$) plane. Symbols denote different amounts of GuHCl, as specified in the legend. Lines are a guide to the eye. Inset: Estimate of the critical temperature $T_\text{c}$ for three different additive concentrations. 
 (B) The same binodals as in (A), but in the normalized second virial coefficient $b_2$ vs. volume fraction $\phi$ plane.
 Data of (A) and (B) are replotted from ref.~\citenum{Hansen2022b}.  }
  \label{fig:f1}
\end{figure}

LLPS can occur when the protein--protein interactions are sufficiently attractive.
Lysozyme in brine represents a model system dominated by short-range attractions,\cite{Sedgwick2005} which can be tuned by additives.\cite{Hansen2016} 
For this system, 
the LLPS binodals and the underlying protein--protein interactions in terms of $b_2$ have been determined previously.\cite{Hansen2022b}
The binodals are shifted by the additives (Fig.~\ref{fig:f1}(A)), but they collapse onto a master curve if the temperature axis is replaced by the second virial coefficient (Fig.~\ref{fig:f1}(B)).

In order to explore the applicability of the ELCS to static and dynamic properties of concentrated protein solutions, in the present work, this model system is investigated 
comprehensively by SLS and DLS experiments.
The system is studied as a function of protein concentration, temperature and additive content in the vicinity of the LLPS binodal.
However, an extremely close proximity as well as highly concentrated samples are avoided; 
therefore, it can be expected that simple, analytical models are applicable to
the $\phi$ dependence of structural and dynamical data 
and mean-field values of the critical exponents describe the asymptotic $T$ dependence of the data.

First, spinodal temperatures are inferred from the $T$ dependence of the SLS data and added to the $b_2$ vs. $\phi$ plane (Fig.~\ref{fig:f1}(A)). 
Then, the $\phi$ dependence of the isothermal compressibility $\kappa_\text{T}$ is analyzed and a corresponding-states behavior is observed.
Third, the relaxation rate $\Gamma$ of the concentration fluctuations is determined and its $T$ dependence is analyzed. 
Fourth, the $\phi$ dependence of the collective diffusion coefficient $D_c$ is analyzed and a corresponding-states behavior is found.

\subsection{Temperature dependence of the isothermal compressibility: corresponding-states behavior of the spinodal}\label{sec:Tkap}

The spinodal line is defined as the limit of thermodynamic stability. 
SLS has previously been used to determine the spinodal line of protein solutions.\cite{Thomson1987,Petsev2003,Manno2003,Manno2004,Bucciarelli2015}
Upon approaching the spinodal by lowering $T$, $\kappa_\text{T}$ is expected to increase according to an asymptotic power law behavior:
\begin{eqnarray}\label{eq:kapT}
\kappa_\mathrm{T} = \kappa_0 \varepsilon^{-\gamma} 
\end{eqnarray}
with a constant $\kappa_0$, the critical exponent $\gamma=1$ in the mean field and the reduced spinodal temperature
\begin{eqnarray}
\varepsilon = \frac{T}{T_\text{s}} -1 
\end{eqnarray}
with the spinodal temperature $T_\text{s}$.
As $\kappa_\text{T}$ diverges at $T_\text{s}$,  $\kappa_\text{T}^{-1}(T_\text{s})=0$.

\begin{figure}[h!]
\includegraphics[width=\columnwidth]{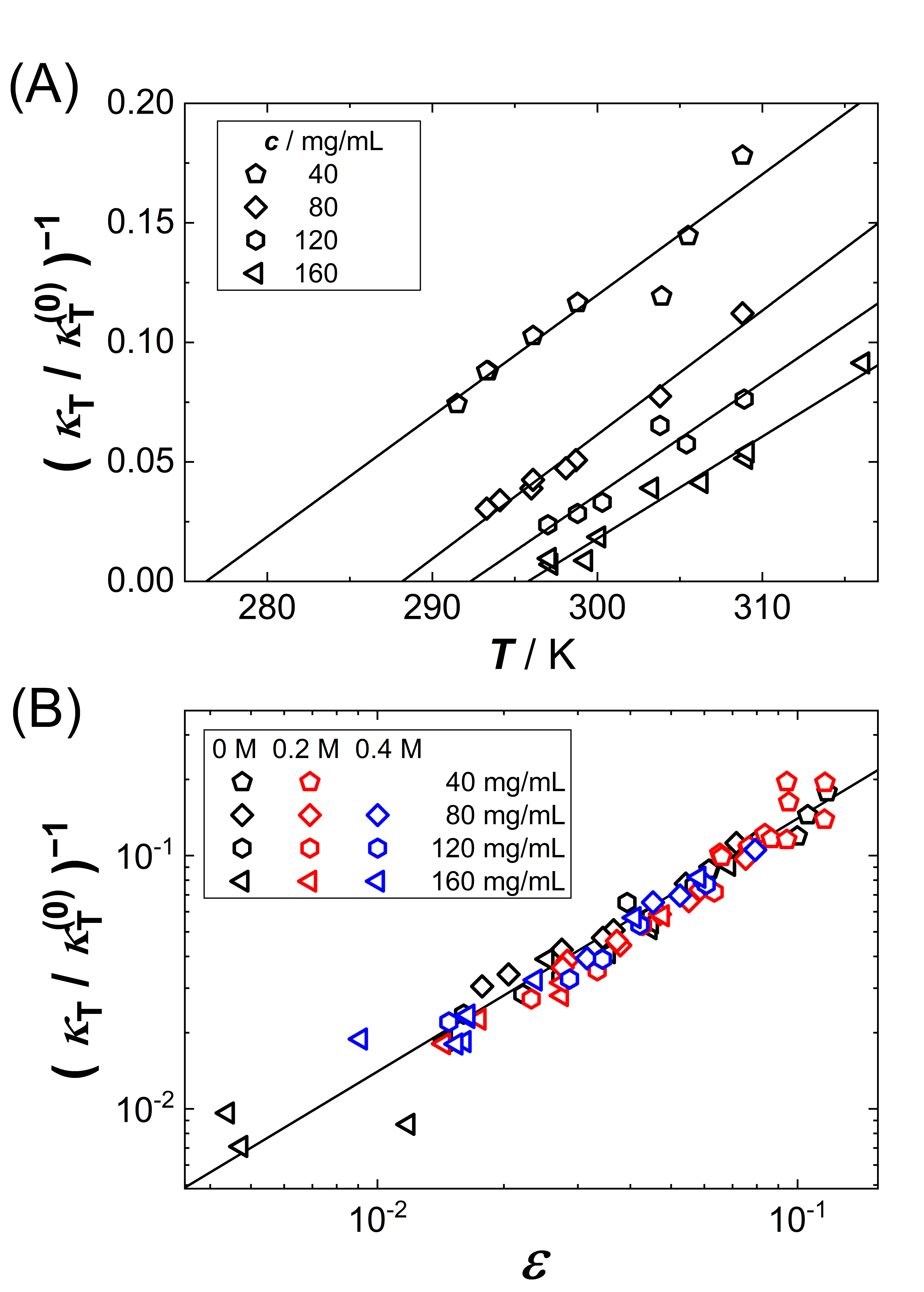}
  \caption{(A) Temperature $T$ dependence of the inverse isothermal compressibility $\kappa_\text{T}$, normalized by $\kappa_\text{T}^{(0)}=1~(\mathrm{N/m}^2)/(\text{mg/mL)}$, for different protein concentrations as indicated: experimental data (symbols) and linear fits (lines).
  (B) Dependence of the inverse isothermal compressibility $\kappa_\text{T}$, normalized by $\kappa_\text{T}^0=1~(\mathrm{N/m}^2)/(\text{mg/mL)}$, on the reduced spinodal temperature $\varepsilon = T/T_\text{s}-1$ where $T_\text{s}$ is the spinodal temperature. Data (symbols), used to determine $T_\text{s}$ shown in (A), i.e. for different GuHCl contents and protein concentrations (columns and rows in the legend, respectively). The solid line has a slope of 1. 
 }
  \label{fig:f2}
\end{figure}

Fig.~\ref{fig:f2}(A) shows exemplary data (symbols) of $\kappa_\text{T}^{-1}(T)$ for various protein concentrations $c$ as indicated.
As expected, the data can be described by linear fits (lines), yielding estimates of $T_\text{s}$ as intercepts of the abscissa.
Note that the spinodal line is submerged below the binodal and hence the $\kappa_\text{T}^{-1}(T)$ data have to be extrapolated by the fits.
Moreover, \Q{at fixed $T$,} $\kappa_\text{T}^{-1}(T)$ decreases with $c$, and hence \Q{larger intercepts of the abscissa and, correspondingly,} $T_\text{s}$ are observed, similar to the \Q{$c$ dependence of the} cloud-point \Q{temperatures} shown in Fig.~\ref{fig:f1}(A).

Fig.~\ref{fig:f2}(B) shows all $\kappa_\text{T}^{-1}$ data as a function of the reduced spinodal temperature $\varepsilon$ together with a solid line of slope 1 on a double logarithmic representation.
As expected from Eq.~(\ref{eq:kapT}), the different data sets collapse onto a single curve which shows a power-law behavior with the mean-field value $\gamma=1$.
The observed scaling behavior further supports the appropriateness of our approach to estimate $T_\text{s}$ values.
\Q{Eq.~(\ref{eq:kapT}) was also fitted globally to the  $\kappa_\text{T}^{-1}$ data with a free, but global value of $\gamma$ (not shown). This procedure yields $\gamma = 0.97$, further supporting the appropriateness of the mean-field value.}

\begin{figure}[h!]
\includegraphics[width=\columnwidth]{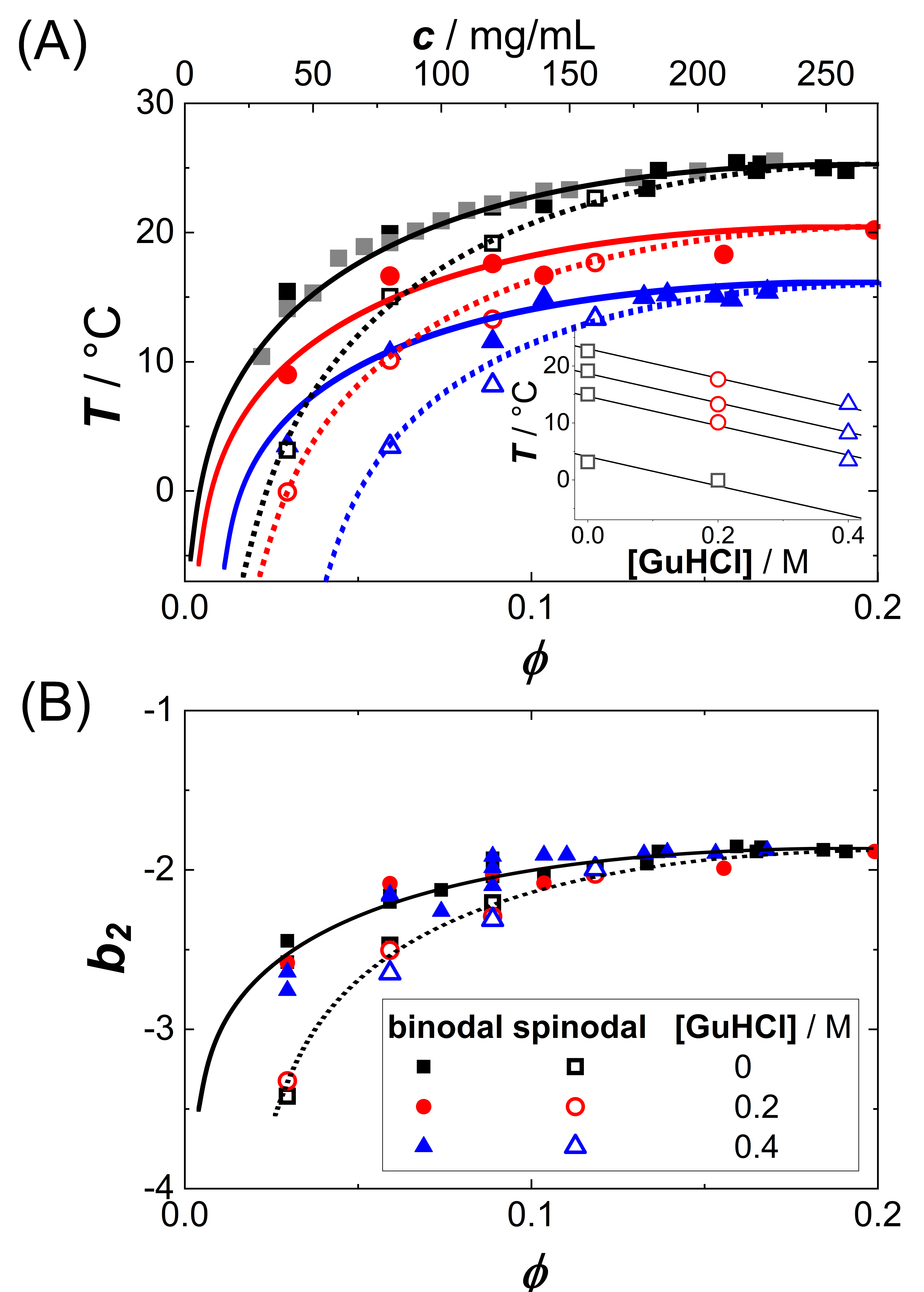} 
  \caption{(A) Binodals (data as full symbols and guides to the eye as solid lines, replotted from Fig.~\ref{fig:f1}, and spinodals (data (open symbols) and eye guides (dotted lines) of lysozyme solutions (0.9~M NaCl) with various amounts of GuHCl as in Fig.~\ref{fig:f1}.
  Inset: Dependence of the spinodal temperatures on the guanidine concentration
  for different protein concentrations (increasing from bottom to top): data (open symbols) and global linear fits (lines).
  (B) The same binodals as in (A), but in the normalized second virial coefficient $b_2$ vs. volume fraction $\phi$ plane. Lines are guides to the eye.   }
  \label{fig:f3}
\end{figure}

In addition to the data obtained in Fig.~\ref{fig:f2}(A), similar experiments have been performed for the two other solution conditions (red circles and blue triangles in Fig.~\ref{fig:f1}(A)). Fig.~\ref{fig:f3}(A) shows the resulting $T_\text{s}$ data (open symbols) and replots the binodals (full symbols).
Indeed, for each of the three conditions, the spinodal is hidden below the binodal and thus also narrower than the binodal. 
With increasing $\phi$, the spinodal approaches the binodal, as they are expected to coincide at the critical point. 
With increasing guanidine content, the spinodal shifts to lower $T$, similar to the decrease of the binodal.
The inset shows the guanidine dependence of $T_\text{s}$ for different protein concentrations together with a common linear fit.
The slope of $-26~\text{K/M}$ agrees with the previously observed value for the binodals.\cite{Platten2015b}
(For one particular solution condition, $c=40~\text{mg/ml}$ and 0.4~M GuHCl, the $\kappa_\text{T}(T)$ are extrapolated over more than $30~\text{K}$, resulting in a very low value of $T_\text{s}$ with large uncertainty. This data point is omitted from further analysis.)

Since $b_2(T)$ data are available for our system (cf. circles in Fig.~\ref{fig:f4}(B)),\cite{Hansen2022b} the spinodal lines of Fig.~\ref{fig:f3}(A) can also be represented in the $b_2$ vs. $\phi$ plane.
Fig.~\ref{fig:f3}(B) shows both the binodal and spinodal lines in this representation.
Again, similar to the binodals, the different spinodal lines collapse onto one another, indicating that the corresponding-states behavior previously observed for the binodals\cite{Platten2015,Hansen2022b} also holds for the spinodals.

\begin{figure}[h!]
\includegraphics[width=\columnwidth]{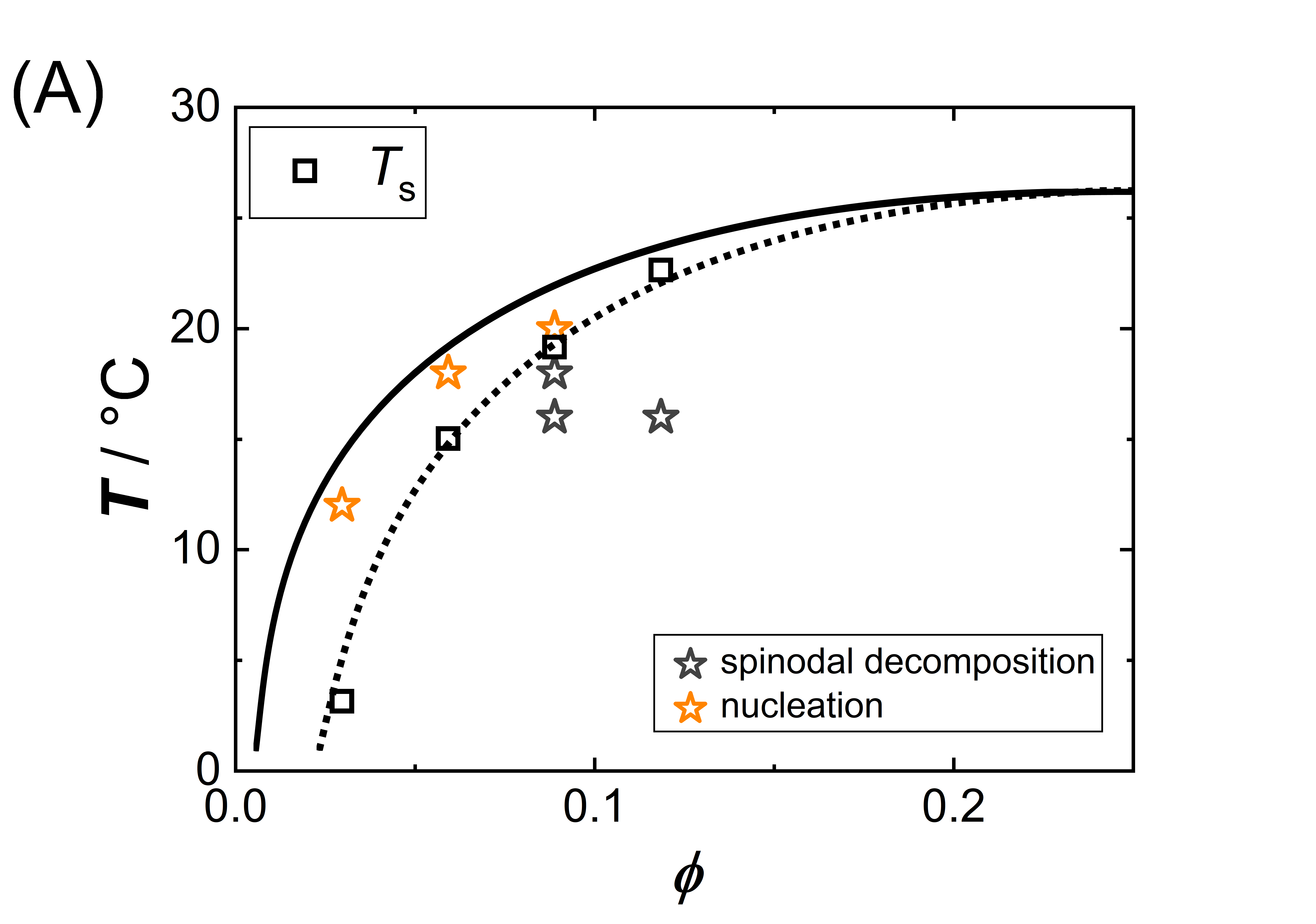}
\includegraphics[width=\columnwidth]{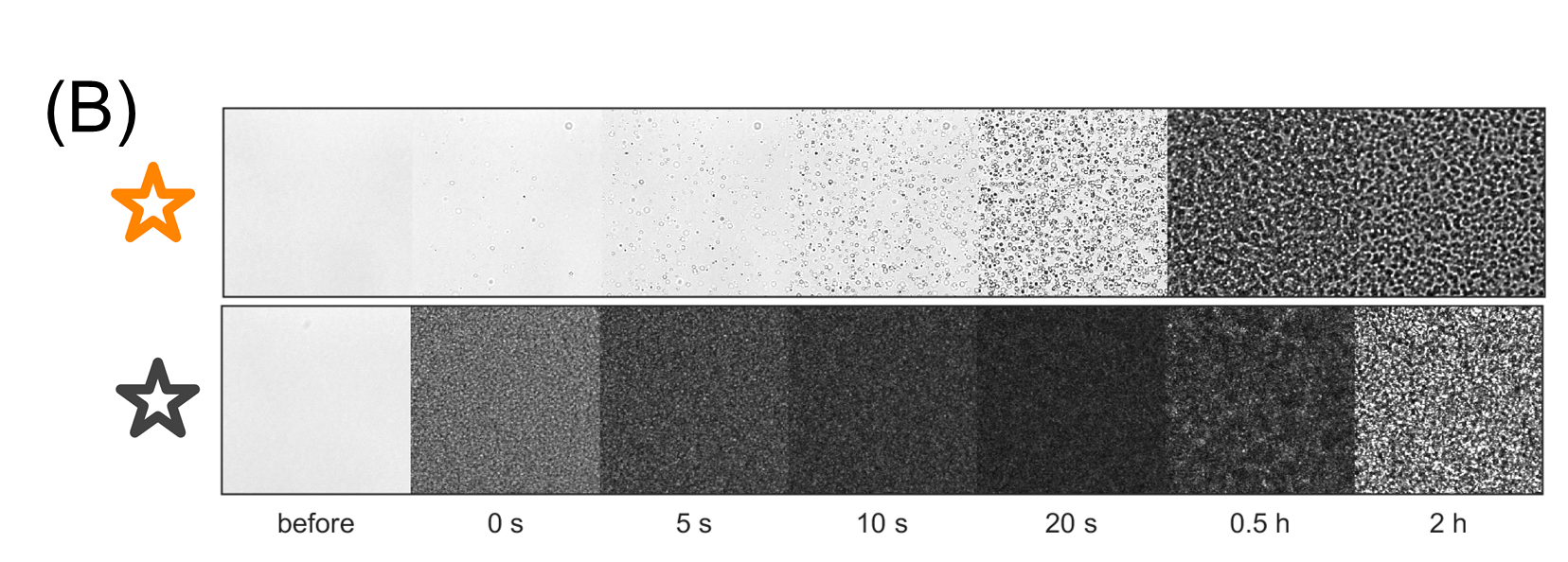} 
  \caption{ (A) Interpolated binodal (solid line), replotted from Fig.~\ref{fig:f1}; spinodal temperatures as inferred from the intercept of the abscissa in Fig.~\ref{fig:f2}(A) connected by a dotted line as a guide to the eye; the solution conditions probed by optical microscopy are indicated by stars.
  (B)   Optical micrographs (image size 248 x 248 $\upmu$m$^2$) illustrating the phase transition kinetics of samples phase separating via nucleation (top) and spinodal decomposition (bottom); in this case, $\phi=0.06$ (top) and $\phi=0.09$ (bottom)  both at $T=18~^\circ\mathrm{C}$, respectively.  }
  \label{fig:kin}
\end{figure}

Some solution conditions (marked by stars in Fig.~\ref{fig:kin}(A)) have been examined by optical microscopy in order to study the phase separation kinetics in the metastable region between spinodal and binodal line as well as in the unstable region below the spinodal.

Exemplary time series of micrographs are shown in Fig.~\ref{fig:kin}(B).
In the metastable region, the microscopy data show the successive formation and growth of droplets, whereas in the unstable region the micrographs reveal a rapid initial darkening as the sample becomes turbid and at later stages the micrographs indicate domain formation and coarsening. 
\Q{The micrographs thus show qualitatively different phase separation kinetics depending on the location of the state. In the metastable region between binodal and spinodal,} phase separation proceeds via droplet nucleation, \Q{while in the unstable region below the spinodal,} spinodal decomposition is observed.

\subsection{Volume fraction dependence of the isothermal compressibility: corresponding-states behavior of $\kappa_\text{T}$}\label{sec:phikap}

\begin{figure}[h!]
\includegraphics[width=\columnwidth]{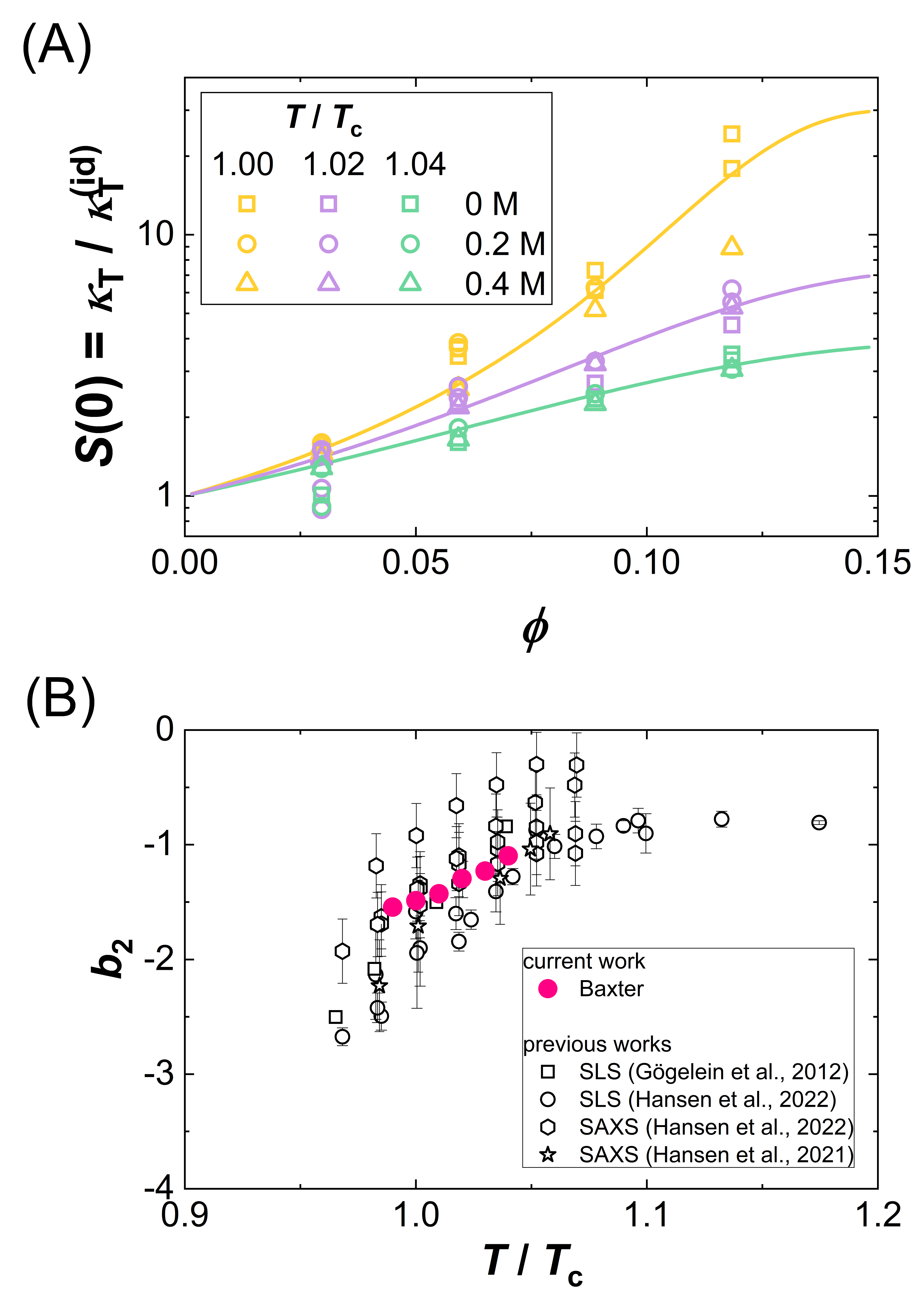} 
  \caption{
  (A) Volume fraction $\phi$ dependence of the isothermal compressibility $\kappa_\text{T}$ normalized by the isothermal compressibility $\kappa_\text{T}^{(\text{id})}$ of an ideal solution: Experimental data (symbols) for different temperatures normalized by the critical temperature, $T/T_\text{c}$, and guanidine concentrations are marked by symbol color and type, respectively. Lines are one-parameter fits to Eq.~(\ref{eq:bax}).
  (B) Normalized second virial coefficient $b_2$ as a function of the reduced temperature $T/T_\text{c}$: Data retrieved from the fits in (A) and literature data\cite{Goegelein2012,Hansen2022b,Hansen2022,Hansen2021b} are shown as full and open symbols, respectively. 
  }
  \label{fig:f4}
\end{figure}

Fig.~\ref{fig:f4}(A) presents exemplarily data on the $\phi$ dependence of $\kappa_\text{T}$, normalized by the respective value of an ideal solution $\kappa_\text{T}^{(\text{id})}$. 
This normalization corresponds to the low-$Q$ limit of the static structure factor or to a normalized derivative of the osmotic pressure and is thus directly related to the osmotic EOS.
Data for different additive 
concentrations [GuHCl] and temperatures $T$, where
the temperature is given relative to the critical temperature, $T/T_\text{c}$, and displayed as symbols in the same color.
Remarkably, though obtained under different solution conditions (i.e., different guanidine concentrations and different absolute temperatures), normalized $\kappa_\text{T}$ data cluster together on a single curve for a given $T/T_\text{c}$.
This collapse indicates a universal osmotic EOS close to $T_\text{c}$ and thus provides an experimental validation of the ELCS.

The $\phi$ dependence of the isothermal compressibility for low salt content\cite{Piazza1998}, i.e., far away from LLPS, has been successfully described by an analytical expression for the compressibility of the adhesive hard-sphere model in the Percus-Yevick approximation\cite{Barboy1974,Regnaut1989,Menon1991,Menon1991b,Chen,Santos}:
\begin{eqnarray}\label{eq:bax}
\frac{\kappa_\mathrm{T}}{\kappa_\mathrm{T}^\mathrm{id}} = \frac{(1-\phi)^4}{(1+2\phi-\lambda_\text{B} \phi)^2}
\end{eqnarray}
with the parameter
\begin{eqnarray}
\lambda_\text{B} = 6 \left( 1-\tau + \frac{\tau}{\phi} \right) \, \left( 1- \left( 1 - \frac{1+2/\phi}{6 (1-\tau + \tau/\phi)^2} \right)^{1/2} \right) \,.
\end{eqnarray}
Thus, within this model, $\kappa_\mathrm{T}$ only depends on $\phi$ and on the stickiness $\tau$.
\Q{Using $\tau$ as a free} parameter, \Q{this model is fitted to the data close to LLPS shown in Fig.}~\ref{fig:f4}(A).
The resulting model fits (lines) to Eq.~(\ref{eq:bax}) reproduce the $\phi$ dependence of the experimental data for the different $T/T_\text{c}$.
The fitting parameter, $\tau$, can be expressed in terms of the normalized second virial coefficient $b_2$ via
\begin{eqnarray}
b_2 = 1-\frac{1}{4\tau} \, .
\end{eqnarray}
Fig.~\ref{fig:f4}(B) shows the resulting values (full symbols), as retrieved from model fits to all $\kappa_\mathrm{T}$ data of the present work.
The magnitude of $b_2$ is found to increase with $T/T_\text{c}$, i.e., attractions are weakened for temperatures further away from the binodal.
The value observed for $T/T_\text{c} \approx 1$ is close to $b_2 = -1.5$, as proposed by Vliegenthart and Lekkerkerker\cite{Vliegenthart2000}.
Moreover, the figure contains literature data\cite{Goegelein2012,Hansen2022b,Hansen2022,Hansen2021b} on $b_2(T/T_\text{c})$ for the present system obtained by light scattering from dilute solutions and from the analysis of small-angle X-ray scattering experiments.
The present data and the independent literature data agree with each other, supporting \Q{the significance of the obtained model parameter ($\tau$) as well as} the appropriateness of the model.
\Q{Further support stems from independent SAXS measurements, whose $Q$ dependent structure factor contribution was accurately described by approximate, analytical expression of the Baxter model.\cite{Hansen2022}}

\subsection{Temperature dependence of the relaxation rate of concentration fluctuations: off-critical slowing down correlated with divergence of the compressibility}

In the preceding sections~\ref{sec:Tkap} and \ref{sec:phikap}, structural properties of protein solutions are analyzed, whereas dynamical properties are investigated in this and the following section.
$\kappa_\text{T}$ quantifies the amplitude of concentration fluctuations\cite{Stanley1971} and shows an asymptotic power-law behavior when $T$ becomes closer to $T_\text{s}$. 
Correspondingly, the relaxation time  (or the inverse relaxation rate $\Gamma^{-1}$) as well as the correlation length of the fluctuations $\xi$ are also expected to diverge.\cite{Stanley1971}
In addition to $\kappa_\text{T}$, $\Gamma$ was determined by DLS experiments on the same solutions.

\begin{figure}[h!]
\includegraphics[width=\columnwidth]{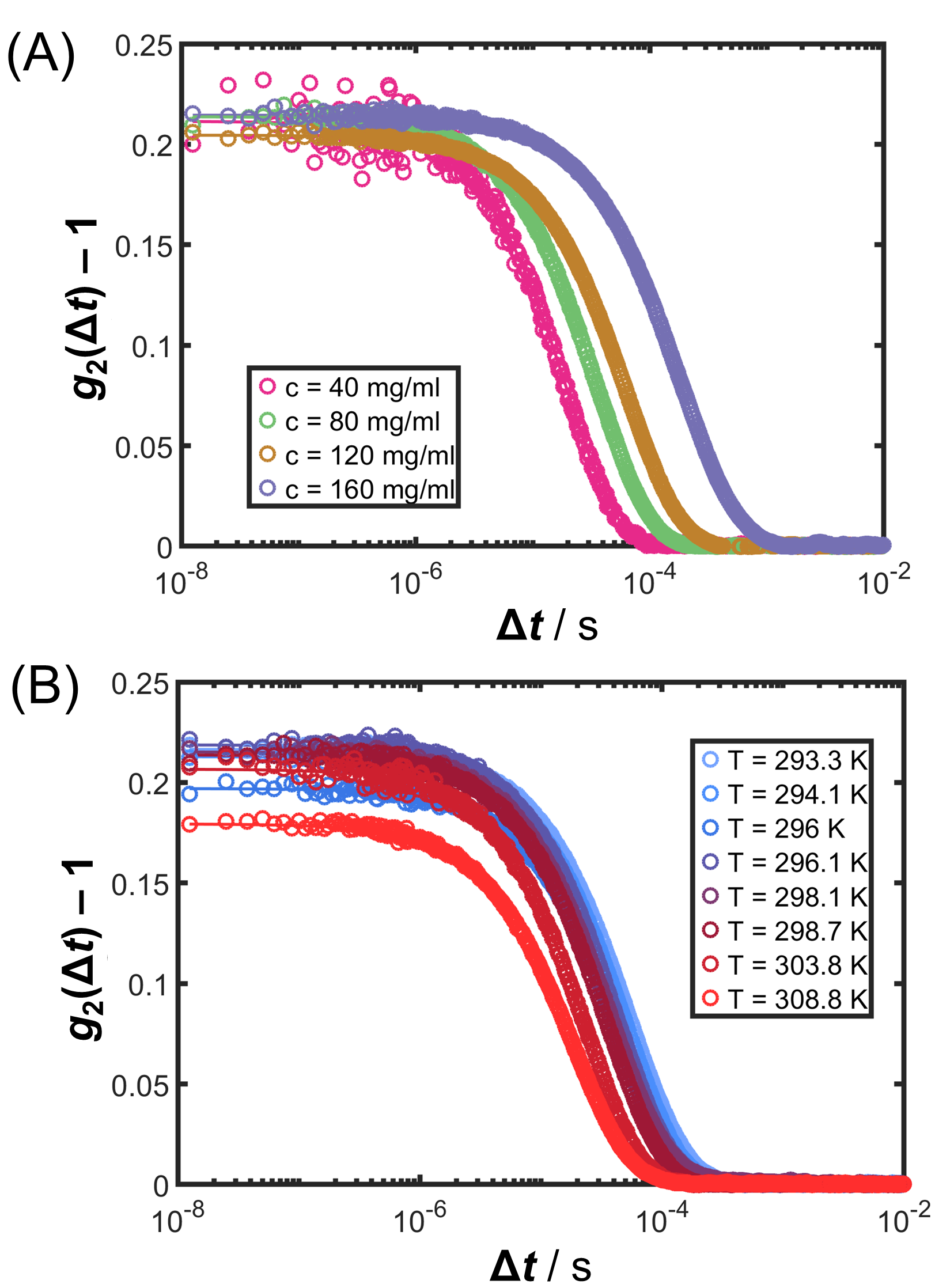} 
  \caption{Intensity cross-correlation functions $g_2(\Delta t)$ as a function of the lag time $\Delta t$ (A) for different protein concentrations (as indicated) at $T=26~^\circ\mathrm{C}$ and (B) for different $T$ (as indicated) at $c=80~\mathrm{mg/mL}$: data (symbols) and second-order cumulant fits (lines).
  }
  \label{fig:f5}
\end{figure}

Fig.~\ref{fig:f5} shows exemplary intensity cross-correlation functions $g_2(\Delta t)$ as symbols (A) at fixed $T=26~^\circ\text{C}$ for different $c$ and (B) at fixed $c=80~\text{mg/ml}$ for different $T$ (as indicated).
All ISFs exhibit a single exponential decay as characteristic for ergodic liquids.
\Q{The ISFs can hence be accurately described by the second-order cumulant ansatz in Eq.~(\ref{eq:cum}).
The corresponding fits (lines) agree with the data.}
\Q{Moreover, the single exponential decay further indicates the absence of aggregates or large impurities and hence also validates the analysis of the SLS experiments.}

It is important to note that most previous DLS studies on the relaxation rate of concentration fluctuations of proteins close to LLPS\cite{Tanaka1977,Ishimoto1977,Manno2003,Manno2004} 
did not take into account possible effects of multiple scattering; some works\cite{Fine1995,Fine1996} reported non-exponential correlation functions.
In the present work, contributions from multiple scattering are suppressed by the 3D cross-correlation set-up and typical single exponential decays of the ISF are observed.
\Q{In arrested systems, stretched, non-exponential decays are expected,\cite{Bandyopadhay2006,Madsen2010} whereas the (off-)critical slowing down rather results in exponential decays.\cite{Manno2004,Bucciarelli2015}}

\begin{figure}[h!]
\includegraphics[width=\columnwidth]{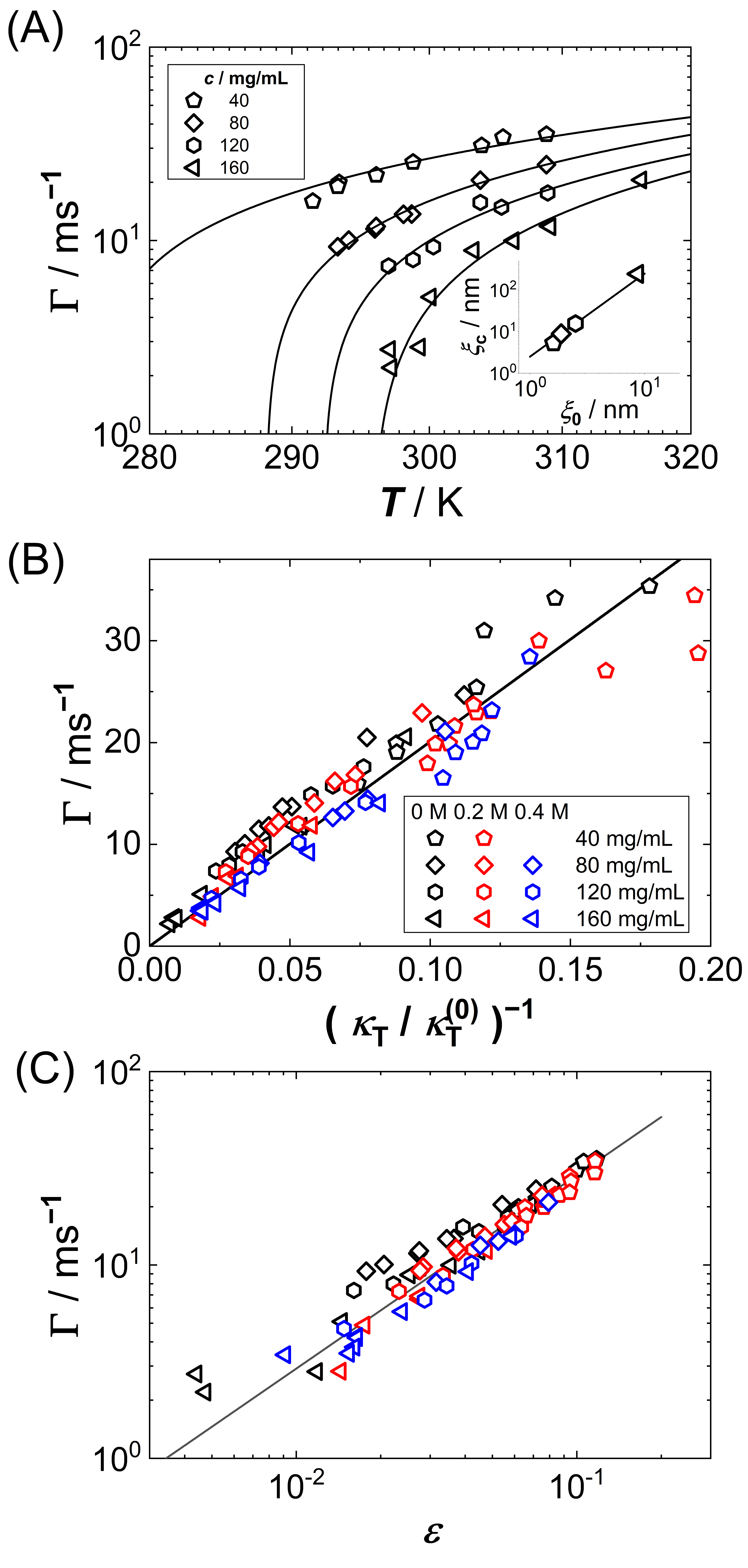} 
  \caption{
  (A) $T$ dependence of the relaxation rate $\Gamma$
  of lysozyme solutions (0.9~M NaCl, no GuHCl added) with protein concentrations as indicated: experimental data (symbols) and fits to Eq.~(\ref{eq:mode}) (lines).
  Inset: Dependence of the correlation length $\xi_\text{c}$ on the correlation length $\xi_0$. Both parameters (symbols) are retrieved from the fits. The line has a slope of 2.
  (B) Dependence of the relaxation rate $\Gamma$ on the inverse isothermal compressibility $\kappa_\text{T}$, normalized by $\kappa_\text{T}^0=1~(\mathrm{N/m}^2)/(\text{mg/mL)}$, for various guanidine and protein concentrations indicated as columns and rows in the caption.
  (C) Dependence of the relaxation rate $\Gamma$ on the 
  reduced spinodal temperature $\varepsilon = T/T_\text{s}-1$ with the spinodal temperature $T_\text{s}$. Symbols as in (B). The solid line has a slope of 1.
  }
  \label{fig:f6}
\end{figure}

Fig.~\ref{fig:f6}(A) shows the $T$ dependence of $\Gamma$ for different $c$, as retrieved from fits of Eq.(\ref{eq:sieg}) and (\ref{eq:cum}) to ISFs.
As expected from the inspection of the ISFs in Fig.~\ref{fig:f5}, $\Gamma$ decreases with increasing $c$ or decreasing $T$, i.e. the concentration fluctuations exhibit a slowing down upon approaching $T_\text{s}$.

In order to quantitatively understand the observed slowing down, 
$\Gamma$ is plotted as a function of $\kappa_\text{T}^{-1}$ in Fig.~\ref{fig:f6}(B). 
A remarkably linear correlation is observed.
As in almost all of our experiments, $\varepsilon$ is not very small \Q{$(\varepsilon>0.01)$} and hence a mean-field picture might be applicable. Then,
the relaxation rate $\Gamma$ and the compressibility $\kappa_\text{T}$ can be related via:\cite{Swinney1973,Berne1976}
\begin{eqnarray}\label{eq:ons}
\Gamma = \frac{\alpha Q^2}{\rho \kappa_{\text{T}}}
\end{eqnarray}
with the number density $\rho$ and the Onsager coefficient $\alpha$, which can be expected to only weakly depend on $T$ and the viscosity $\eta$.
Eq.~(\ref{eq:ons}) together with Eq.~(\ref{eq:kapT}) thus explains the observed $c$ and $T$ dependence of $\Gamma$, where the latter appears governed by the diverging behavior of $\kappa_\text{T}$.
If $\alpha$ is assumed to assumed to be constant in the relevant range, Eq.~(\ref{eq:kapT}) implies: $\Gamma\sim\varepsilon^\gamma$ with $\gamma=1$ in the mean field.
Fig.~\ref{fig:f6}(C) shows the dependence of $\Gamma$ on the reduced spinodal temperature $\varepsilon$. 
Indeed, a linear correlation is observed as expected from Fig.~\ref{fig:f2}(B) and \ref{fig:f6}(B).
\Q{A global power-law fit to the $\Gamma(\varepsilon)$ data (not shown) with a common, but variable $\gamma$ yielded $\gamma = 0.96 \pm 0.04$, further supporting the appropriateness of the mean-field approach.}
This is in contrast to previous findings\cite{Manno2003}, where an exponent of $0.63$ was observed for similar solution conditions, but contributions from multiple scattering were not suppressed.

In a more general mode-coupling framework,\cite{Kawasaki1970,Oxtoby1974,Burstyn1982}
the relaxation rate can be described as the sum of a critical and a background contribution in the vicinity of the critical point (or the spinodal).
If the correlation length is small compared to $Q^{-1}$, as \Q{for most of our data}, the relaxation rate has a simple form:\cite{Sorensen1988}
\begin{eqnarray}\label{eq:mode}
\Gamma = \frac{k_\text{B} T Q^2}{6 \pi \eta_0} \frac{1}{\xi} \left( \frac{T_\text{s}}{T} \frac{\xi_c}{\xi} + 1 \right)
\end{eqnarray}
with the correlation length of the concentration fluctuations
given by
\begin{eqnarray}\label{eq:xi}
\xi = \xi_0 \varepsilon^{-\nu}
\end{eqnarray}
with a constant $\xi_0$ and a critical exponent $\nu$ (with a mean-field value $\nu=0.5$), a constant viscosity $\eta_0$ and a parameter $\xi_\text{c}$ expected to scale as $\xi_0^2$.
A two-parameter fit ($\xi_0$ and $\xi_c$) of Eq.~(\ref{eq:mode}) is performed to the data shown in Fig.~\ref{fig:f6}(A).
The fits (lines) quantitatively reproduce the data and yield estimates of the correlation length $\xi$ according to Eq.~(\ref{eq:xi}) which are shown in the inset of Fig.~\ref{fig:f6}(A).

\subsection{Volume fraction dependence of the relaxation rate of concentration fluctuations: corresponding-states behavior of the collective diffusion coefficient}

In order to relate the dynamics to the underlying interactions, it is more convenient to express the relaxation rate of concentration fluctuations in terms of the collective diffusion coefficient $D_\text{c}$.
When comparing various solution conditions (additive compositions and temperatures) with each other, it is moreover useful to normalize $D_\text{c}$ by the diffusion coefficient of non-interacting single particles, $D_0$.
Its value can be obtained from the low $c$ limit of $D_\text{c}$ or it can be estimated based on the Stokes-Einstein relation:
\begin{eqnarray}\label{eq:SE}
D_0=\frac{k_\text{B} T }{6 \pi \eta(T) R_\text{H}} \,.
\end{eqnarray}
with the viscosity of the solution, $\eta(T)$, estimated based on literature data\cite{Kestin1981,Kawahara1966}, 
and the apparent hydrodynamic radius $R_\text{H}=2.2~\text{nm}$. This value is chosen slightly larger than previously reported,\cite{Muschol1995} which might be due to the large ion content and account for uncertainties in $\eta$.

\begin{figure}[h!]
\includegraphics[width=\columnwidth]{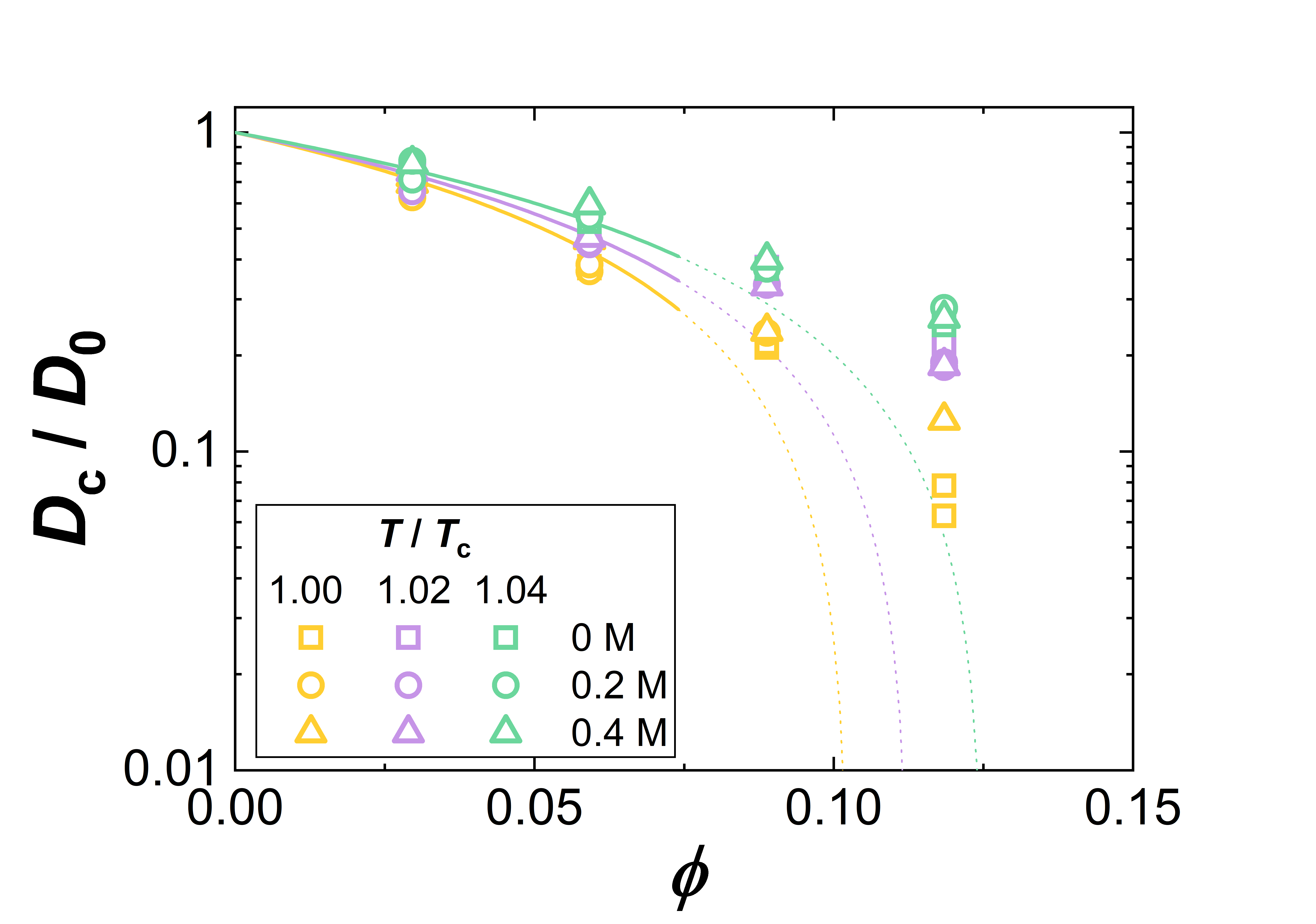} 
  \caption{Volume fraction
  $\phi$ dependence of the collective diffusion coefficient of concentration fluctuations $D_\text{c}$ normalized by the diffusion coefficient at infinite dilution $D_0$. 
  Experimental data for various reduced temperatures $T/T_\text{c}$ and guanidine concentrations are marked by symbol color and type, respectively. 
  Lines are computed based on Eq.~(\ref{eq:cicho}) using $\tau$ as retrieved from the fits shown in Fig.~\ref{fig:f4}.
  }
  \label{fig:f7}
\end{figure}

Fig.~\ref{fig:f7} shows exemplarily data on the $\phi$ dependence of $D_\text{c} /D_0$. 
As in Fig.~\ref{fig:f4}(A), data for different additive compositions, but the same temperature relative to the critical temperature, $T/T_\text{c}$, are displayed as symbols in the same color.
Again, as for $\kappa_\text{T}/\kappa_\text{T}^\text{(id)}$, though obtained under different solution conditions, the
$D_\text{c} /D_0$ data collapse onto a single curve for fixed $T/T_\text{c}$.
This indicates a corresponding-states behavior of the collective diffusion coefficient close to $T_\text{c}$.
Hence, our data provide further experimental support of the ELCS, also for dynamical properties.

As for compressibility data, the simple Baxter model can be applied to analyze the collective diffusion.
An approximate equation for $D_\text{c} (\phi) /D_0$, to first order in $\phi$,\cite{Cichocki1990} is available:
\begin{eqnarray}\label{eq:cicho}
\frac{D_\text{c}}{D_0} = 1+\lambda_\text{c}\phi + \text{O}(\phi^2)
\end{eqnarray}
with
\begin{eqnarray}
\lambda_\text{c} = 1.454 - \frac{1.125}{\tau} \, .
\end{eqnarray}
Thus Eq.~(\ref{eq:cicho}) depends only on the stickiness $\tau$.
This indicates that, based on the same value for $\tau$ used to describe the compressibilities in Fig.~\ref{fig:f4}, 
$D_\text{c}(\phi) /D_0$ can be calculated without any free parameter.
The results (lines) agree with the data at low concentrations (Fig.~\ref{fig:f7}), but as expected deviate at larger $\phi$, where the linear term (shown as dotted line) is insufficient.

\subsection{Discussion}

\Q{In this section, the relevance of our findings and their applicability to other solution conditions and other protein systems is commented on with respect to the applicability of the ELCS and the use of coarse-grained colloidal models.}

\Q{The data presented in Fig~\ref{fig:f4}(A) and Fig.~\ref{fig:f7}  indicate that, close to LLPS, the osmotic compressibility and the collective diffusion coefficient is only determined by the volume fraction and the temperature relative to the critical temperature or, equivalently, by the second virial coefficient, as shown in Fig.~\ref{fig:f4}(B).
Thus, for our system, the ELCS applies and rationalizes static and dynamical properties of concentrated solutions.
Note that, for the solution conditions investigated here, the repulsive interactions are largely screened and very similar for the various conditions. 
However, if one considers systems in which repulsion is more important, it is conceivable that this has to be accounted for in an effective particle size.\cite{Noro2000,Platten2015}
Nevertheless, the present data as well as previous studies\cite{Gibaud2011,Platten2015,Bucciarelli2016,Hansen2022,Hansen2022b} suggest that the ELCS applies to a broad range of proteins and solution conditions.}

\Q{The adhesive hard-sphere model proposed by Baxter\cite{Baxter1968} represents one of the simplest systems with short-range attractions.
It is therefore astonishing that an approximate theoretical description for this model is suitable to quantitatively describe the complex interactions between protein molecules even in the vicinity of LLPS.
While the ELCS suggests that the detailed shape of the interactions does not matter, this does not guarantee that the approximate theoretical description can also be reasonably applied to describe the experimental conditions. 
It is conjectured that the success of the description might also be related to the globular shape of lysozyme, the moderate concentrations considered as well as the not-to-close proximity to the spinodal.
It is conceivable that models accounting for the directionality of the interactions\cite{Goegelein2008,Roosen-Runge2014,Skar2019} might be more appropriate to describe the structural and dynamical properties of concentrated solutions of proteins of more complex shape.
However, as implied by the ELCS, for the conditions analyzed here, similar results are expected.}

\section{Conclusion}

\Q{
The extended law of corresponding states suggests that, close to LLPS, thermodynamical properties of protein solutions are not sensitive to the details of the underlying interactions.
The present work aims at a comprehensive experimental test of the ELCS.
For solution conditions with different protein concentrations, temperatures and additive contents, static and dynamic light scattering experiments were performed.
The isothermal osmotic compressibility of the solutions was determined based on the scattered light intensity. 
From the $T$ dependence of $\kappa_\text{T}$, spinodal temperatures $T_\text{s}$ were estimated and found to collapse onto a master curve if temperatures are expressed in terms of the second virial coefficient as for the binodals.
The intensity cross-correlation function was characterized by a single exponential decay for the conditions studied.
The relaxation rate of the concentration fluctuations was found to be correlated with $\kappa_\text{T}$.
If considered at the same temperature normalized by the critical temperature, values of $\kappa_\text{T}$ collapse onto a master curve, providing an experimental validation of the ELCS.
In addition, a similar collapse is observed for the collective diffusion coefficient.
The $\phi$ dependence of $\kappa_\text{T}$ and $D_\text{c}$ is accurately reproduced by the approximate, analytical expressions for adhesive hard-spheres with interaction parameters that agree with independent data. 
Our results hence support and broaden the applicability of the ELCS to the static and dynamical properties of protein solutions. 
They also rationalize and justify the use of coarse-grained colloidal models for protein solutions close to LLPS.
}

\section*{Conflicts of interest}
There are no conflicts to declare.

\section*{Acknowledgements}
We thank Gerhard N\"{a}gele (FZ Jülich, Germany) and  Ram\'on Casta\~neda-Priego (Leon, Mexico) for stimulating and very helpful discussions.
F.P. acknowledges financial support by the German Research Foundation (PL 869/2-1).



\balance


\bibliography{lorena} 
\bibliographystyle{rsc} 
 

\end{document}